\documentclass[twoside,a4paper,11pt]{proceedings}
\usepackage{graphicx}
\usepackage{hyperref}
\usepackage{movie15}
\usepackage{natbib}
\usepackage{wrapfig}

\newcommand{\Ha}{H$\alpha$}
\newcommand{\Hb}{H$\beta$}
\newcommand{\SII}{[S {\sc ii}]}

\newcommand{\NII}{[N {\sc ii}]}

\newcommand{\OIII}{[O {\sc iii}]}

\newcommand{\FeII}{[Fe {\sc ii}]}

\newcommand{\HII}{H{\sc ii}}

\begin{document}
\pagenumbering{arabic}
\pagestyle{myheadings}
\thispagestyle{empty}
\vspace*{-1cm}
{\flushleft\includegraphics[width=3cm,viewport=0 -30 200 -20]{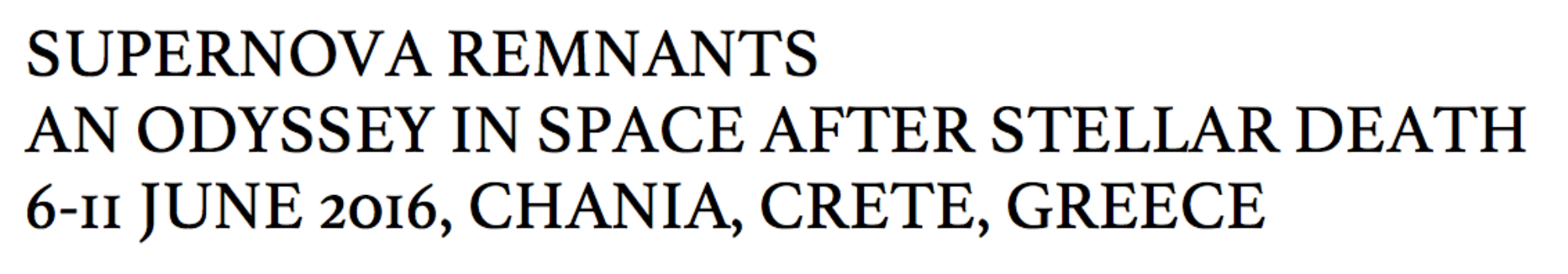}}
\vspace*{0.2cm}
\begin{flushleft}
{\bf {\LARGE
Probing the properties of extragalactic Supernova Remnants
}\\
\vspace*{1cm}
I. Leonidaki $^{1, 2}$
%
}\\
\vspace*{0.5cm}
%

$^{1}$
IESL/Foundation for Research and Technology-Hellas, 71110 Heraklion, Crete, Greece \\
$^{2}$
Department of Physics, University of Crete, GR-71003, Heraklion, Crete, Greece \\


%
\end{flushleft}
\markboth{
Extragalactic SNR populations
}{
I. Leonidaki
}
\thispagestyle{empty}
\vspace*{0.4cm}
\begin{minipage}[l]{0.09\textwidth}
\ 
\end{minipage}
\begin{minipage}[r]{0.9\textwidth}
\vspace{1cm}
\section*{Abstract}{\small

The investigation of extragalactic SNRs gives us the advantage of surmounting the challenges we are usually confronted with when observing Galactic SNRs, most notably Galactic extinction and distance uncertainties. At the same time, by obtaining larger samples of SNRs, we are allowed to cover a wider range of environments and ISM parameters than our Galaxy, providing us a more complete and representative picture of SNR populations. I will outline the recent progress on extragalactic surveys of SNR populations focusing on the optical, radio, and X-ray bands. Multi-wavelength surveys can provide several key aspects of the physical processes taking place during the evolution of SNRs while at the same time can overcome possible selection effects that are inherent from monochromatic surveys. I will discuss the properties derived in each band (e.g. line ratios, luminosities, denisties, temperatures) and their connection in order to yield information on various aspects of their behaviour and evolution. For example their interplay with the surrounding medium, their correlation with star formation activity, their luminosity distributions and their dependence on galaxy types.
\vspace{10mm}
\normalsize}
\end{minipage}

\section{Introduction}



The evolution of Supernova Remnants (SNRs) within a uniform Interstellar Medium (ISM) is well-described by a four-stage model, introduced by \citet{Woltjer1972}: the free expansion phase where the ejecta of the Supernova sweeps up matter as it expands freely until the mass of the ejecta equals the mass of the swept up material. Then it passes to the adiabatic phase at which the SNR's evolution can be described by the Sedov-Taylor self-similar solution (\citealt{Sedov1959}; \citealt{Taylor1950}). The first two phases (Fig.\,1) depict the blast waves of the newly formed SNRs, reaching high shock velocities (5$\times$10$^{3}$ - 10$^{4}$ km/s) and heating the material behind the shock front to temperatures up to 10$^{8}$, producing thermal X-rays (for a good review on X-ray emission in SNRs see \citealt{Vink2012}). The third stage (Fig.\,1) occurs when the mass of the swept up 
\begin{wrapfigure}{r}{0.45\textwidth} 
\vspace{-17pt}
\hspace{-23pt}
\includegraphics[width=0.4\textwidth, angle=-90]{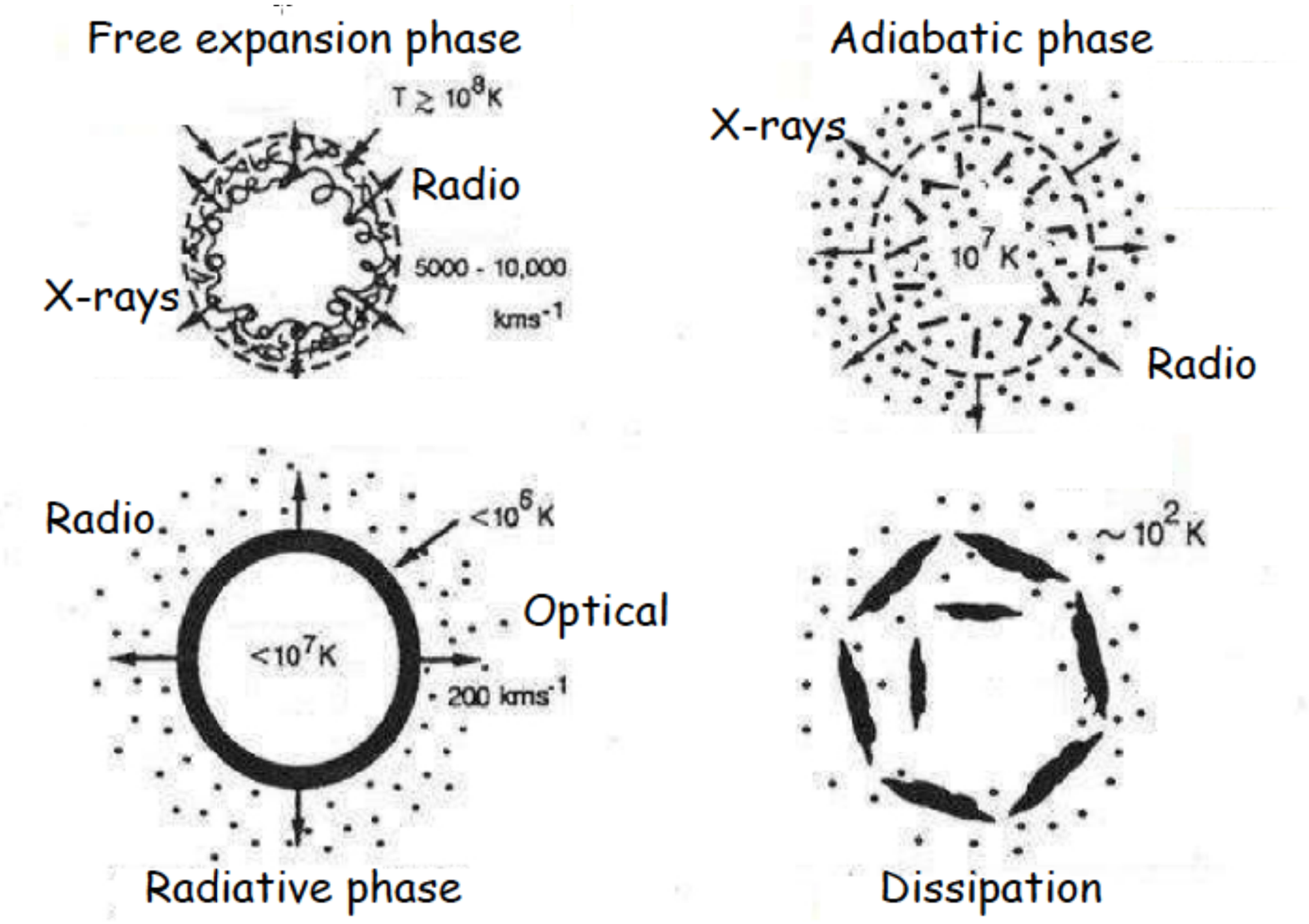}
\vspace{-10pt}
\caption{Evolutionary stages of a Supernova Remnant}
\vspace{-10pt}
\end{wrapfigure} 
material has dramatically increased, forcing the velocity of the shock front to decrease down to $\sim$200 km/s. The temperature behind the shock front drops to $\sim$10$^{5}$ and the energy losses due to recombination become significant, creating a cooling region behind the shock front and producing shock-heated collisionally ionized species (such as \SII, \OIII\, or hydrogen recombination lines). This is the first time the SNR is radiating in the optical band. The final stage of evolution occurs when the velocity of the shock reaches the sound speed of the ambient ISM and the SNR dissipates. Synchrotron radio emission is present throughout the life of the remnant as it is produced mainly at the vicinity of the shock (e.g. \citealt{Charles1995}). From the aforementioned it is evident that to what extend a SNR becomes an X-ray, optical or radio emitter depends on its evolutionary stage.\\ 
Apart from the evolutionary stage/age, the derived properties of SNRs depend on various other parameters such as the environment/ISM (e.g. density, temperature),  the progenitor properties (e.g. mass loss rate, stellar wind density, composition) and selection effects. However, the details on the connection between these properties are poorly understood while each one of these parameters has its own signature at different wavebands: For example, different wavebands can yield information on density and temperature for different gas phases, the SNR progenitors can be evaluated using the X-ray spectra from the ejecta of young SNRs or optical SNRs are easier to be detected in low density/diffuse emission regions than radio or X-ray SNRs. Therefore, it is essential to investigate SNRs in a multi-wavelength context in order to have a more complete picture about their evolution.\\


\subsection{Milky Way vs Extragalactic SNRs: Pros and Cons}

Galactic SNRs allow us to probe the physics from individual regions of the remnants and their interaction with their surrounding ISM. However, these studies are severely hampered by two crucial factors: Galactic absorption and distance uncertainties. Most of the Galactic SNRs are located in the galactic plane impeding the detection of optical or X-ray emitting SNRs while due to distance uncertainties, essential parameters such as sizes or luminosities cannot be estimated. Therefore, there are difficulties in conducting systematic studies or probing their evolution, although an adequate number of Galactic SNRs is in hand (294 sources; \citealt{Green2014}). \\
On the other hand, the study of extragalactic SNRs presents many advantages: they are regarded at the same distance with the observed galaxy, effects of internal Galactic absorption can be minimized (especially when we study face-on galaxies), a wider range of environments and ISM parameters than our Galaxy can be selected (e.g. different metallicities, star formation histories, masses) providing us this way a more complete and representative picture of the SNR populations, while larger samples can be obtained with fewer observations. Although there is limited sensitivity and spatial resolution in these kind of studies, it is imperative to sample large samples of galaxies in order to understand the global properties and the systematics of the SNR populations as a function with their environment.

\begin{figure}
\centering
\vspace{-45pt}
\hspace{-35pt}
\begin{minipage}{.5\textwidth}
\centering
  \includegraphics[width=0.8\linewidth, angle=-90]{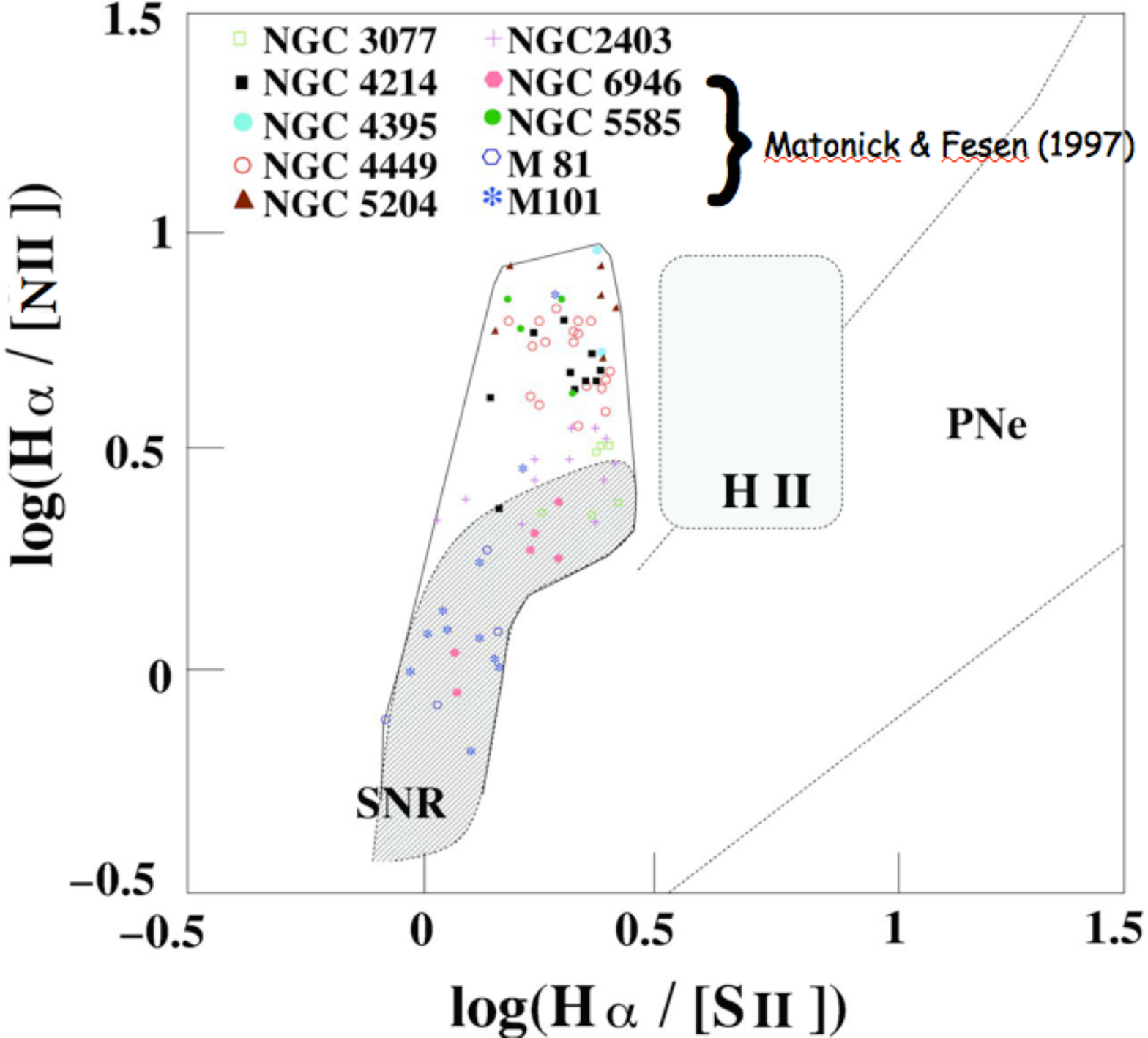}
  \label{fig:test1}
\end{minipage}%
\begin{minipage}{.5\textwidth}
\centering
\includegraphics[width=.93\linewidth]{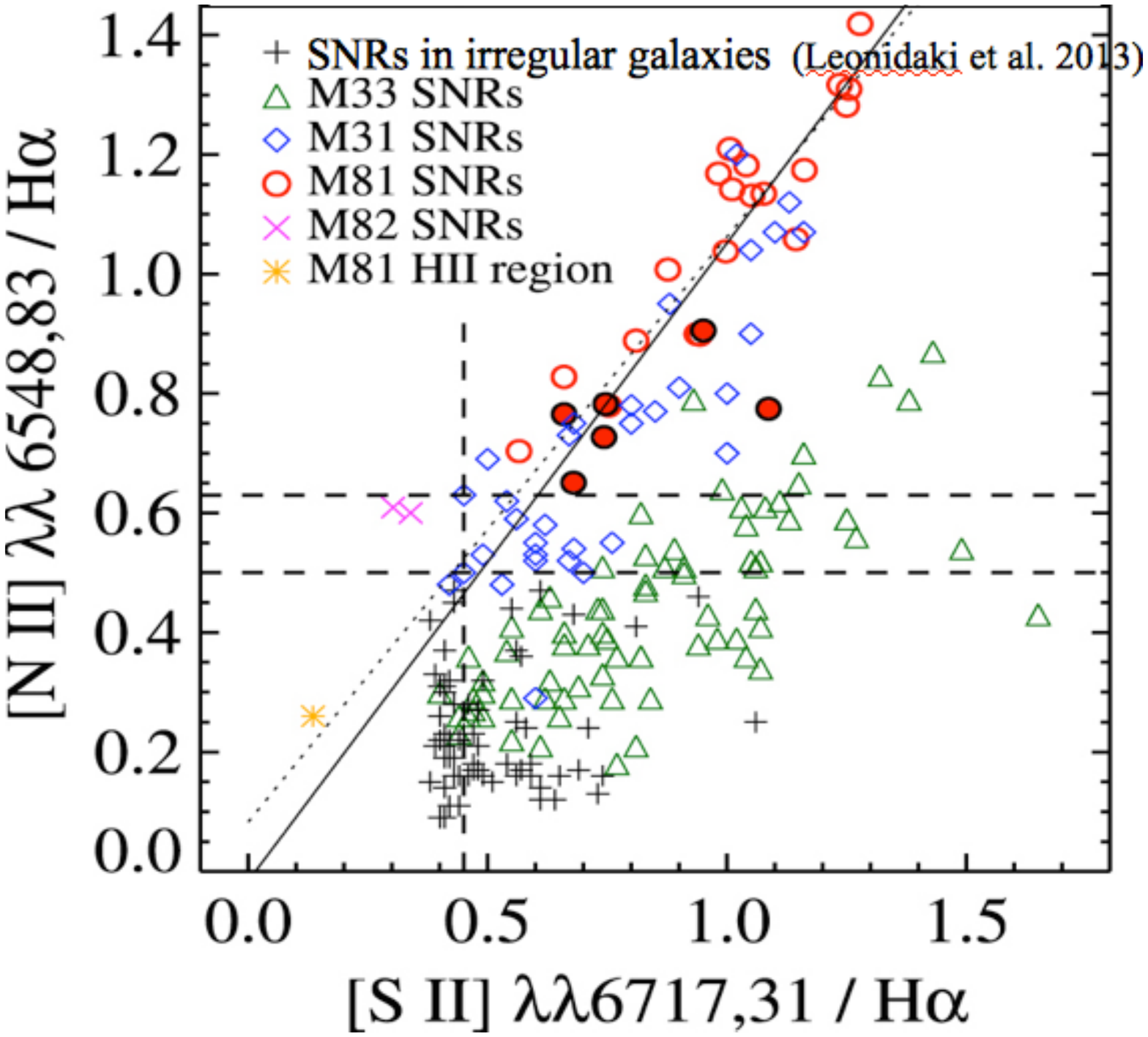}
\vspace{-7pt}
  \label{fig:test2}
\end{minipage}
\vspace{-45pt}
\caption{\SII/\Ha\, - \NII/\Ha\, emission line ratios of spectroscopically observed SNRs in various and different types of galaxies (see \S2.1)}
\vspace{-8pt}
\end{figure}

\vspace{-10pt}
\subsection{Selection criteria for detecting SNRs}

The widely-used selection criterion for SNRs comes from the optical band and is the emission line ratio \SII/\Ha$>$ 0.4 (for Type II SNRs) which has been empirically diagnosted to well-differentiate shock-excited (SNRs) from photo-ionized (\HII\, regions, Planetary Nebulae) regions (\citealt{Mathewson1973}). Radio SNRs can be easily disentangled using their non-thermal synchrotron emission. Thermal X-ray emitting SNRs appear to have temperatures below 2 keV and a thermal plasma spectrum (\citealt{Leonidaki2010}). In the case of non-thermal X-ray emitting SNRs (Pulsar Wind Nebulae - PWN) and since they have very similar X-ray spectra to X-ray Binaries (XRBs), they cannot be identified based solely on their X-ray properties. Another diagnostic used to identify many new remnants (some of them could hardly be located with optical images) is the \FeII\, 1.644 $\mu$m emission line (\citealt{Blair2014}). Nevertheless, to what extend an SNR can be detected in a multi-wavelength context depends on its evolutionary stage and/or its ambient medium.

\section{SNRs in the optical: emission line diagnostics}

Emission lines in optical surveys are a very useful tool in estimating various physical parameters. These lines depend on many factors (e.g. abundances, shock velocities, magnetic parameters, excitation mechanisms) and sophisticated models are needed for accurate calculations. However, there are some emission line ratios that have been proven to be indicative of specific parameters (e.g. \SII/\Ha: main shock-heating gas indicator, \NII/\Ha: metallicity indicator and secondary shock-heating gas indicator; \OIII/\Hb: shock velocity indicator; \SII(6716/6731): electron density indicator).

\begin{figure}
\centering
\hspace{-40pt}
\begin{minipage}{.5\textwidth}
\centering
\includegraphics[width=.69\linewidth, angle=-90]{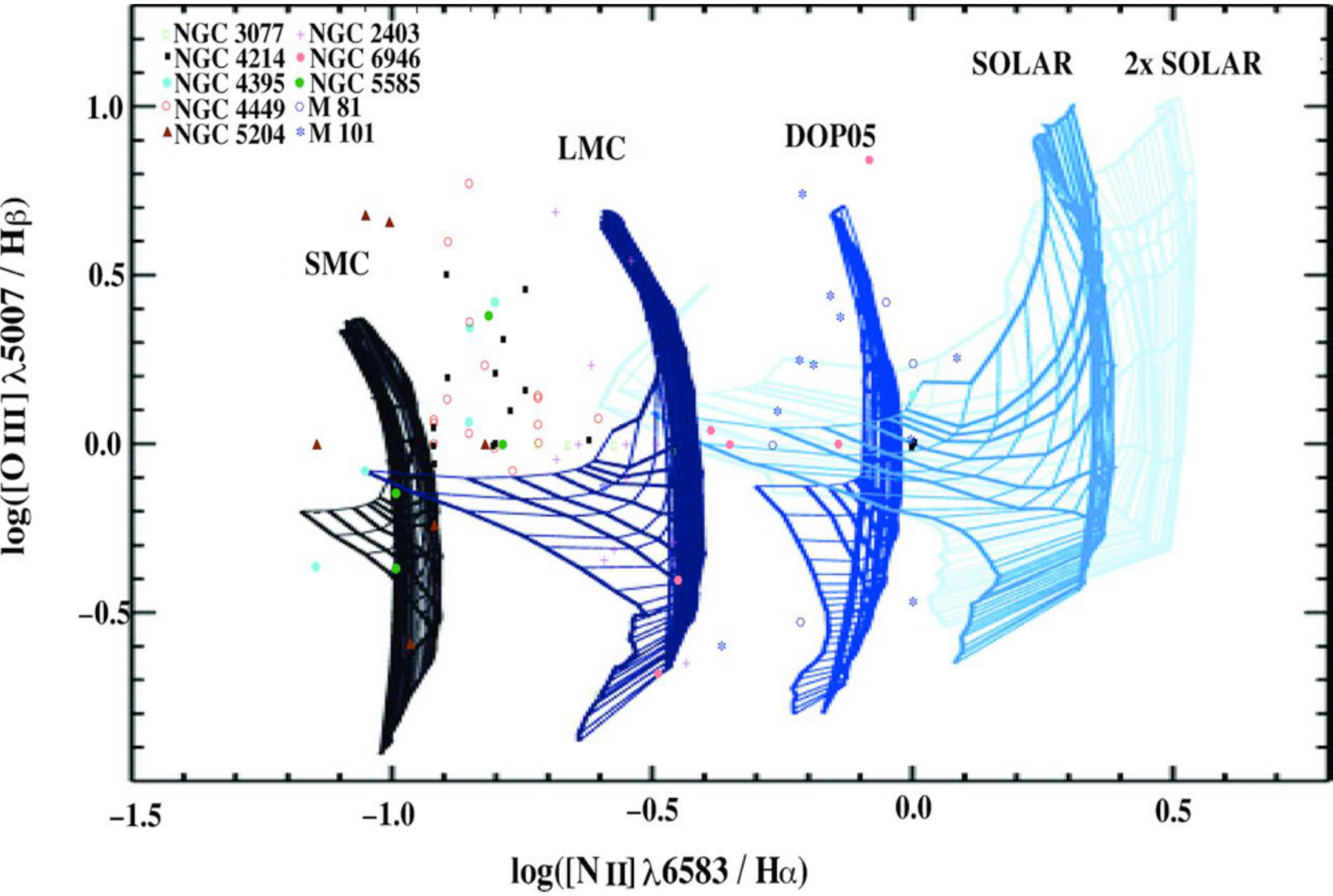}
\label{fig:test1}
\end{minipage}%
\begin{minipage}{.5\textwidth}
\centering
\vspace{-10pt}
\includegraphics[width=.77\linewidth, angle=-90]{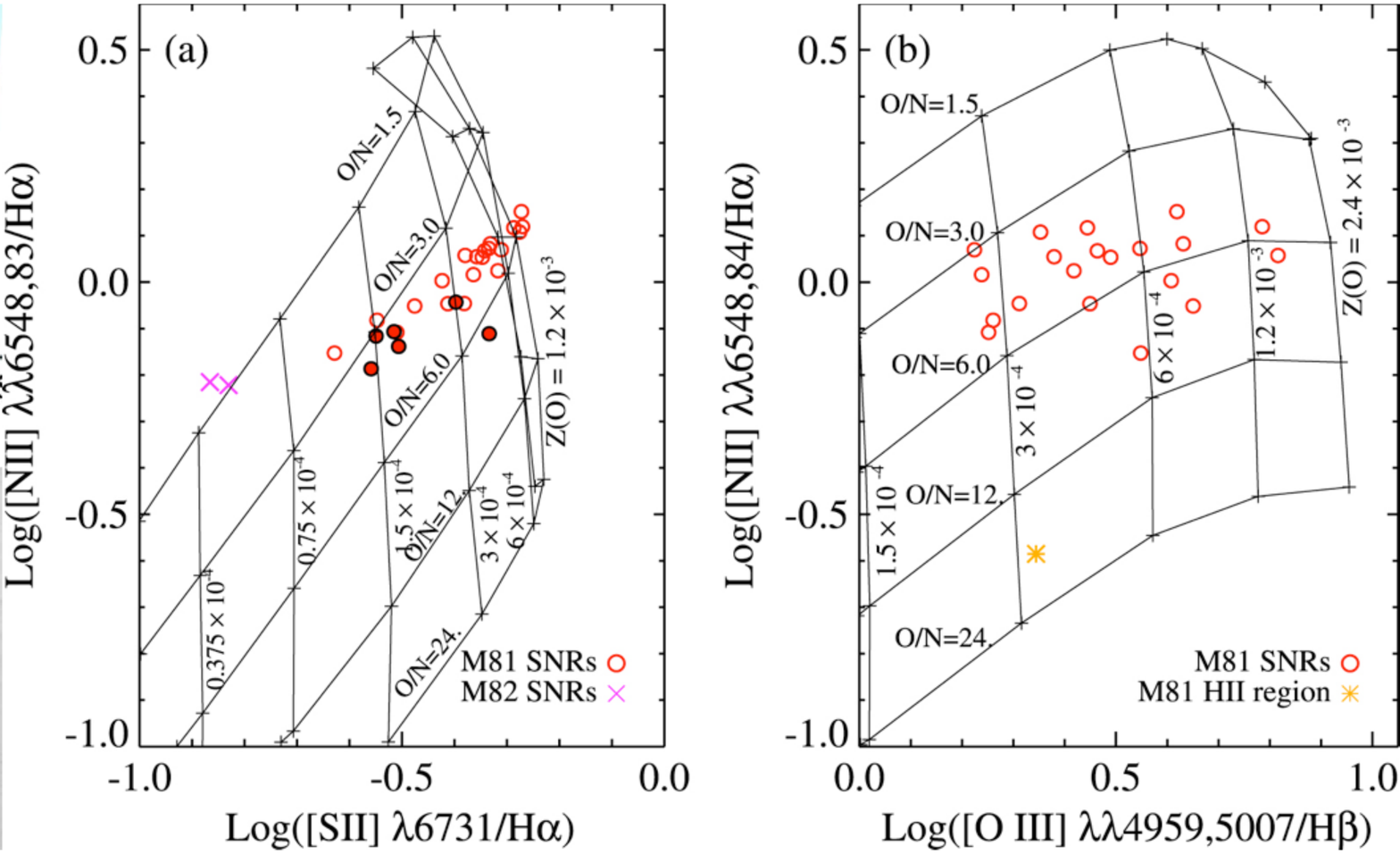}
\vspace{-10pt}
\label{fig:test2}
\end{minipage}
\vspace{-5pt}
\caption{Diagnostic diagrams of \OIII/\Hb\, vs \NII/\Ha\, (left panel; \citealt{Leonidaki2013}) and \NII/\Ha\, vs \SII/\Ha\, or \OIII/\Hb\, (middle, right panels; \citealt{Lee2015}) for estimating shock velocities and abundances respectively}
\end{figure}

\subsection{\NII/\Ha: metallicity and shock-heating gas indicator}

A represenative example of how the \NII/\Ha\, ratio depicts metallicity variations is well shown in Fig.\,2 (left panel), where the log(\Ha/\SII) against the log(\Ha/\NII) emission line ratios of a large number of spectroscopically observed SNRs in six nearby galaxies (5 {\it irregulars} and 1 {\it spiral}) are plotted (\citealt{Leonidaki2013}). For comparison, the spectroscopically observed SNRs of four more {\it spiral} galaxies are included (\citealt{Matonick1997}). The locus of the different types of sources in these diagrams (dashed lines) have been created using the emission line ratios of a large number of Galactic SNRs, \HII\, regions and Planetary Nebulae (PNe) and can help us distinguish the excitation mechanism of the emission lines (photoionization for \HII\, regions and PNe or collisional excitation for SNRs). As can be seen, all sources are within the range of \SII/\Ha\,= 0.4 - 1 which is typical for SNRs. What is intriguing though is that along the \Ha/\NII\, axis, the vast majority of the SNRs in irregular galaxies extend outside the region of Galactic SNRs in contrast to the SNRs of spiral galaxies which occupy that specific region. The region of SNRs in irregular galaxies is shifted in the direction of higher \Ha/\NII\, ratios, indicating weaker emission in the \NII\, lines. This could be either due to difference in excitation or difference in metallicity. However, since there is no particular difference between the \SII/\Ha\, ratios (a powerful shock-excitation indicator for SNRs) for the SNR populations in spiral and irregular galaxies, this suggests a difference in \Ha/\NII\, line ratios of the SNR populations between different types of galaxies due to metallicity. Indeed, irregular galaxies present typically lower metallicities in relation to spiral galaxies (e.g.\citealt{Pagel1981}; \citealt{Garnett2002}).\\
The \NII/\Ha\, ratio can also be used as a good secondary shock-heating gas indicator. Various studies (e.g. \citealt{Blair1997, Blair2004}; \citealt{Lee2015}) have revealed a strong correlation between the \SII/\Ha\, and \NII/\Ha\, ratios of SNRs. This is also evident in Fig.\,2 (right panel) from the work of \citet{Lee2015}: higher \SII/\Ha\, ratios present higher \NII/\Ha\, ratios. What is allso interesting in that plot is that the SNRs in irregular galaxies present a flatter slope (higher \SII/\Ha\, ratios) in comparison to the SNRs in spiral galaxies. This could be interpreted either as differences in metallicity or the presence of non-uniform ISM which is known to be the case in irregular galaxies. The spiral M\,33 is placed in the region of irregulars most probably due to its lower metallicity in relation to the other spirals (see \citealt{Pilyugin2004} for metallicities of numerous galaxies). \\

\begin{figure}
\hspace{+18pt}
\begin{minipage}{.5\textwidth}
\hspace{-70pt}
\vspace{-85pt}
\includegraphics[width=\linewidth, angle=-90]{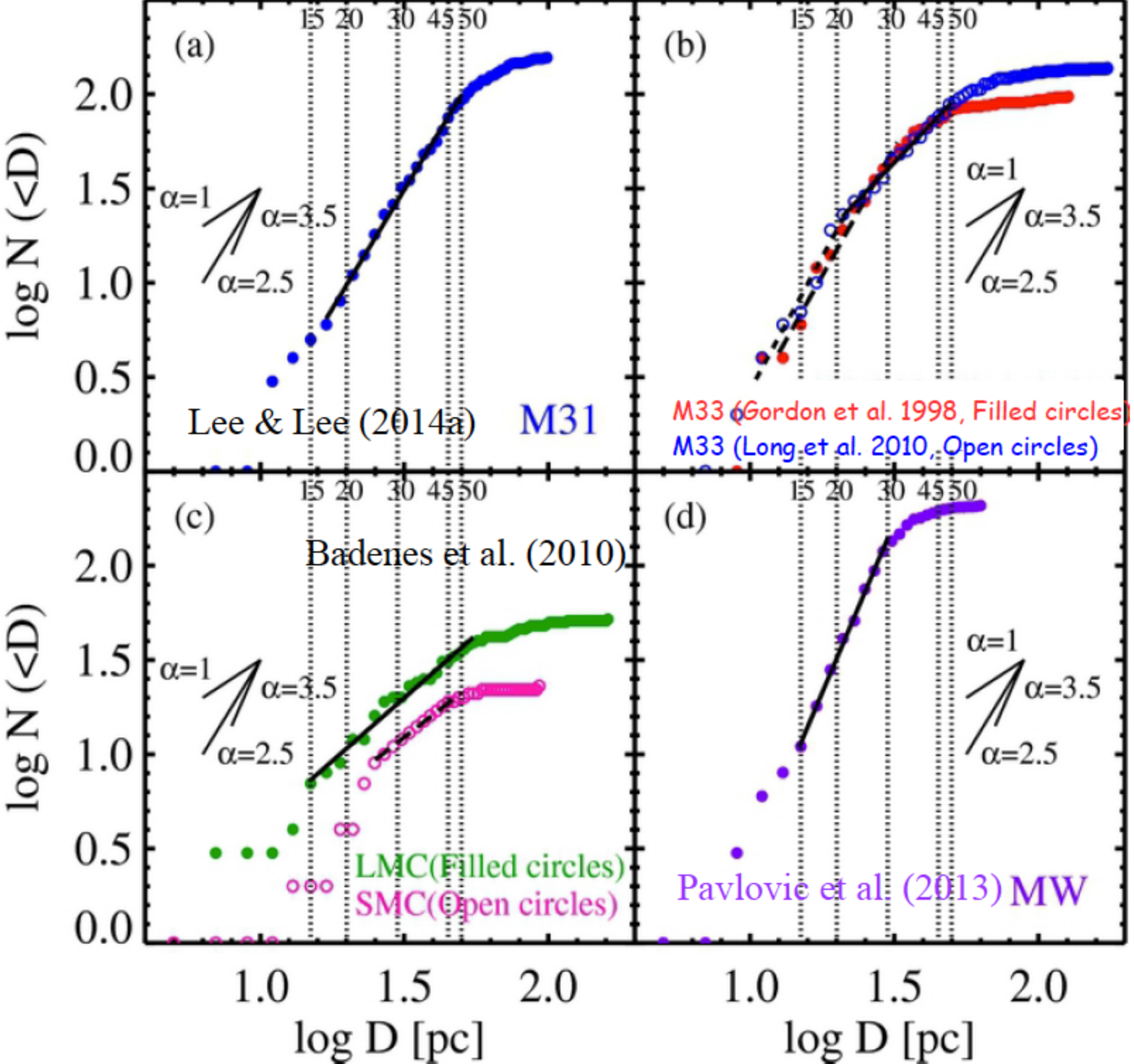}
\vspace{+80pt}
\label{fig:test1}
\end{minipage}
\hspace{+10pt}
\begin{minipage}{.40\textwidth}
\vspace{-25pt}
\centering
\vspace{-22pt}
\includegraphics[width=.9\linewidth]{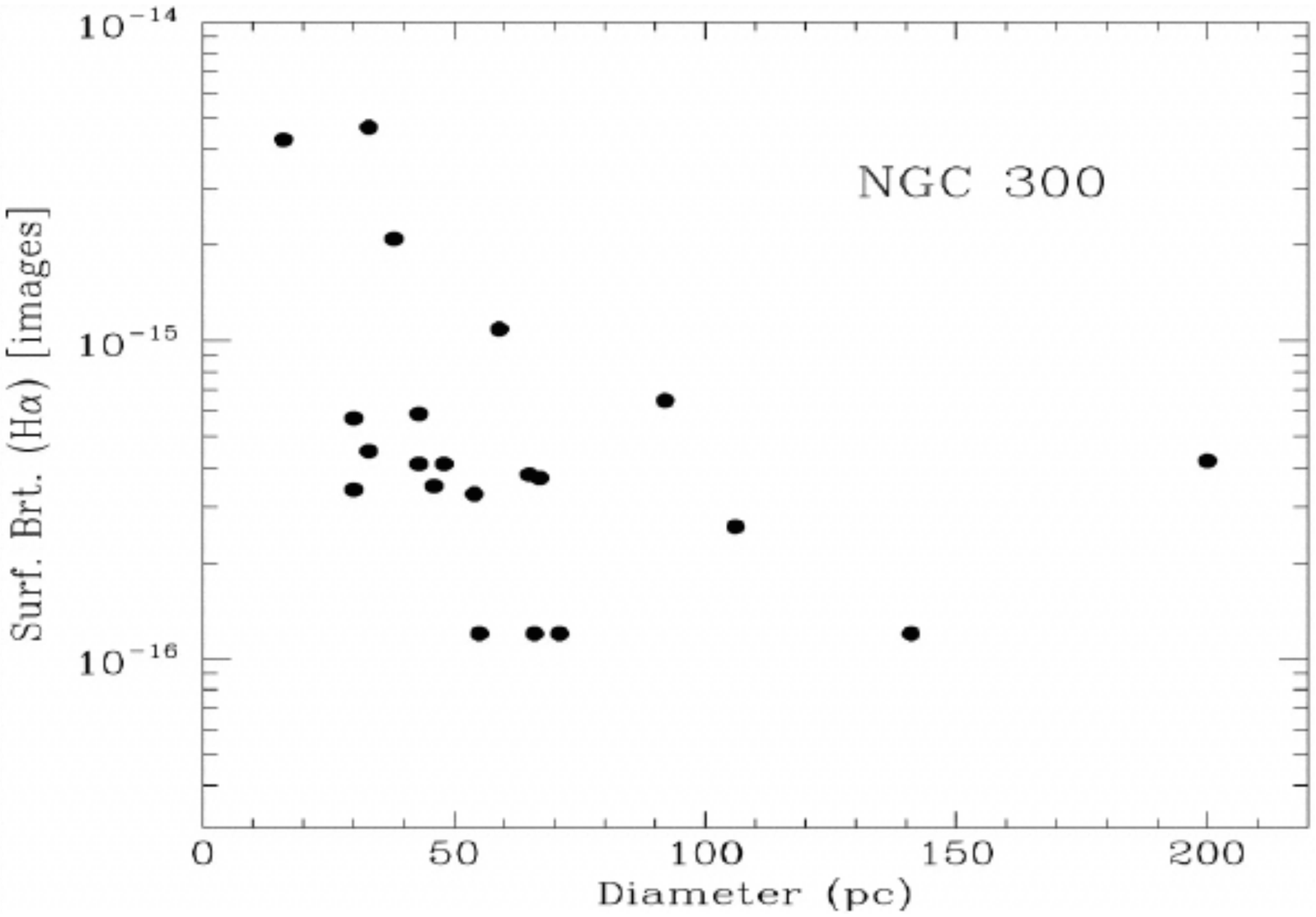}\vspace{-100pt}
\includegraphics[width=.9\linewidth]{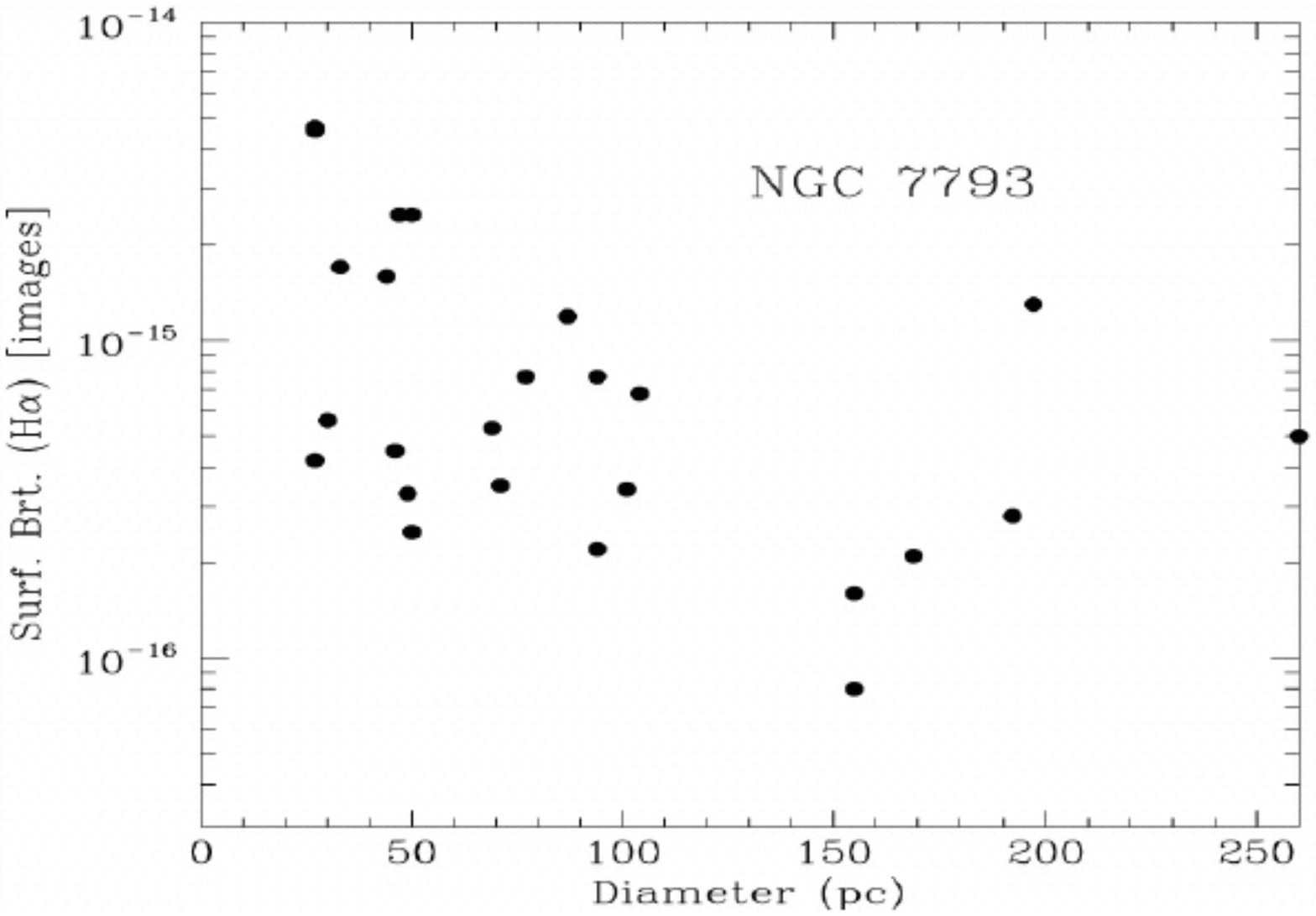}\hspace{-6pt}
\vspace{-50pt}
\label{fig:test2}
\end{minipage}
\caption{{\it Left panel:} Cumulative size distributions of the SNR populations in M\,31, M\,33, the MCs and the Milky Way (\citealt{Lee2014a})(see \S3)  {\it Right panels:} Surface brightness vs diameter of SNRs in NGC\,300 and NGC\,7793 (\citealt{Blair1997})(see \S3)}
\end{figure}

\subsection{Shock models for measuring abundances and shock velocities}

Using sophisticated models we can measure elemental abundances and shock velocities: For example, \citet{Leonidaki2013} used the theoretical shock model grids of \citet{Allen2008} in order to estimate the shock velocities of spectroscopically identified SNRs in six nearby galaxies (Fig.\,3, left panel). These model grids were constructed for different values of shock velocities (horizontal lines starting from 200 km/s on top and going down to 1000 km/s in a step of 50), magnetic field parameters (vertical lines at the grids), and chemical abundances. The \OIII/\Ha, \NII/\Ha\,emission line ratios of the SNRs were plotted on the grids indicating that most of the observed sources have shock velocities less than 200 km/s (indicating evolved SNRs) while only a handful of those present high shock velocities.
Another example comes from the work of \citet{Lee2015} where, using shock ionization models of  \citet{Dopita1984} for various values of oxygen abundance and the ratio of oxygen to nitrogen abundance,  they estimated the oxygen abundances of SNRs in M\,81 (Fig.\,3, middle and right panels).

\begin{figure}
\vspace{-60pt}
\begin{minipage}{.5\textwidth}
\hspace{-20pt}
  \includegraphics[width=0.85\linewidth]{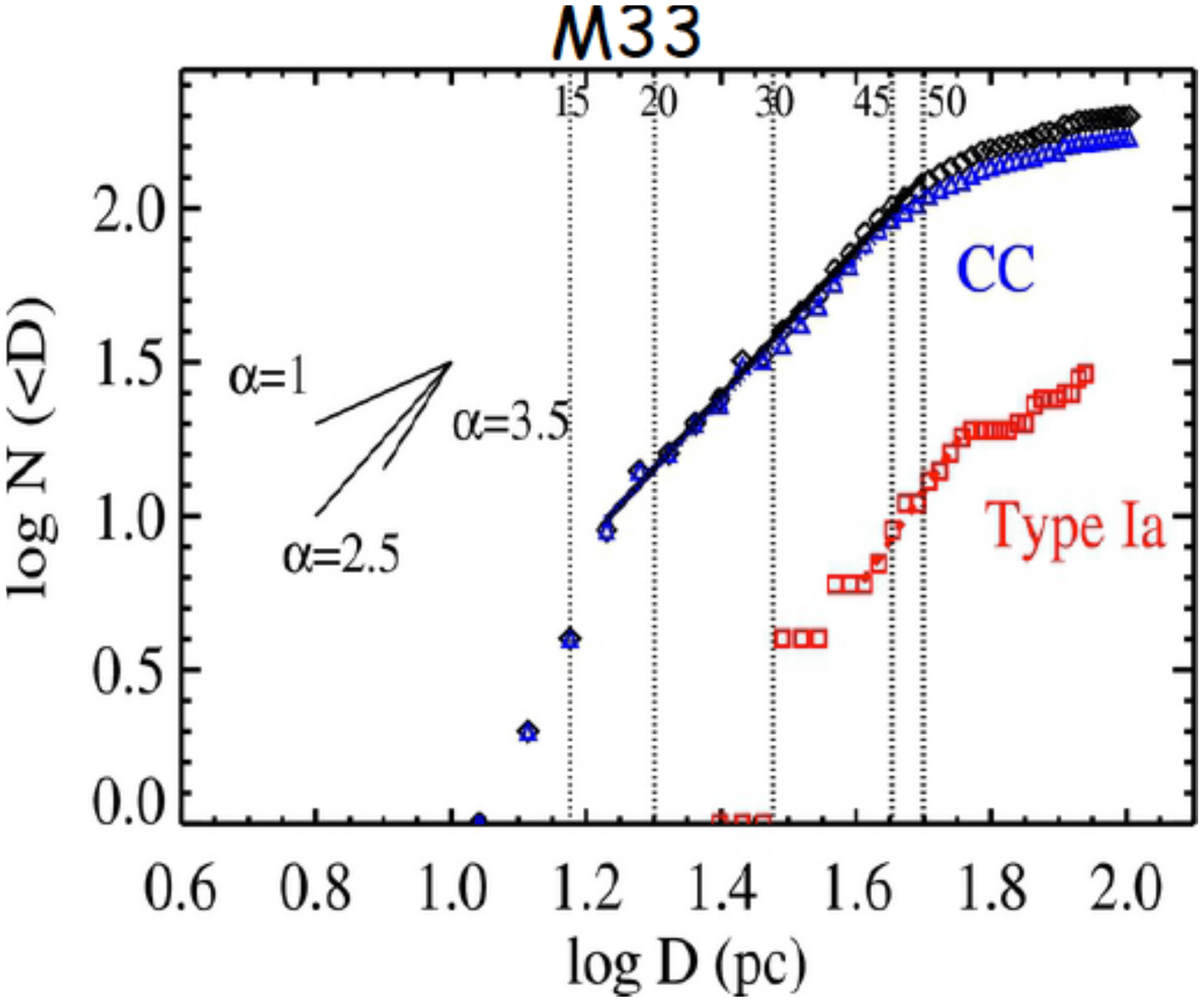}
\end{minipage}%
\begin{minipage}{.5\textwidth}
\hspace{-65pt}
\includegraphics[width=0.98\linewidth, angle=-90]{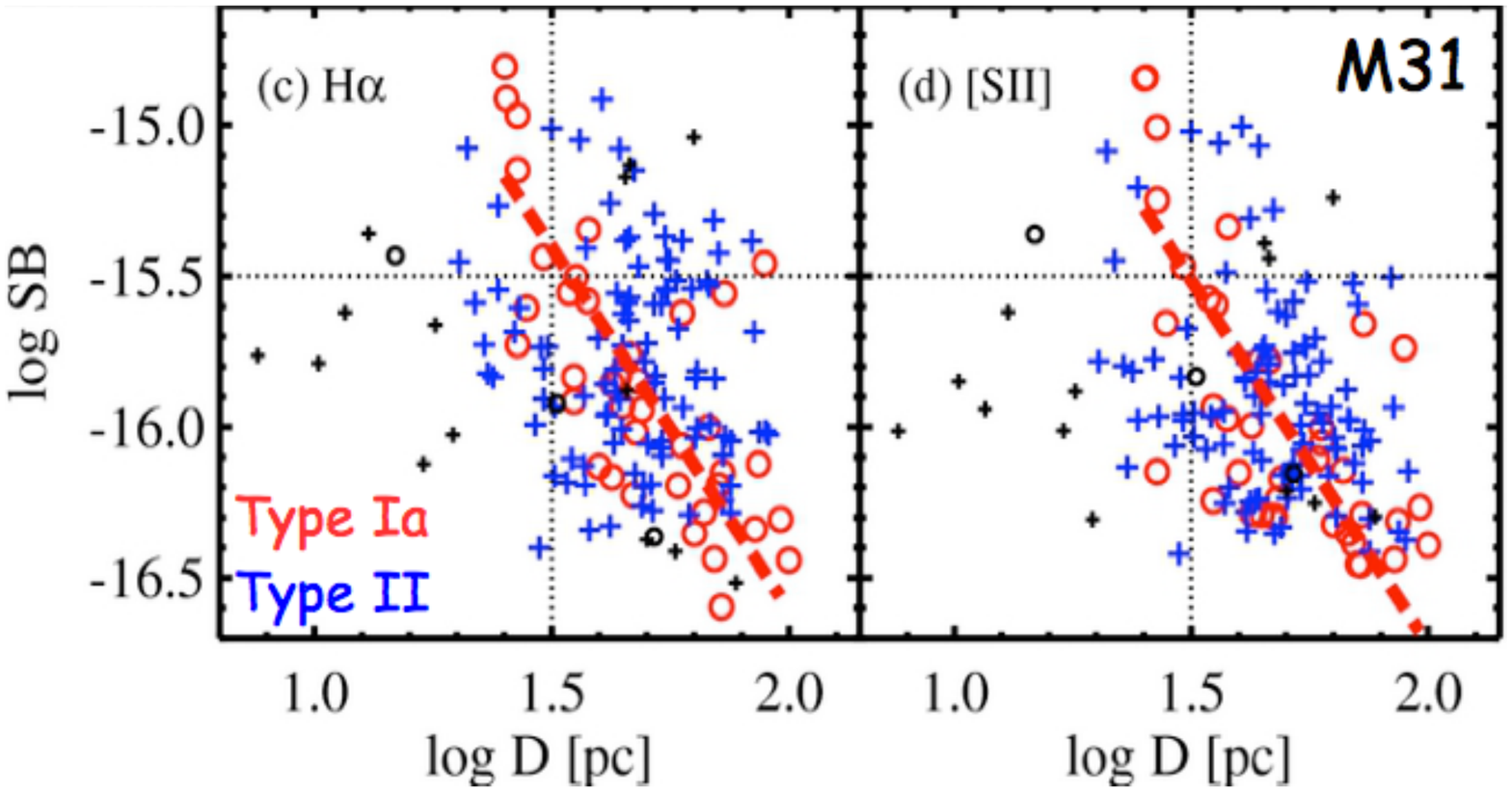}\vspace{-13pt}
\end{minipage}
\vspace{-60pt}
\caption{{\it Left:} Cumulative size distribution of M\,33 (\citealt{Lee2014b}), based on different SNR types (Type II/Ia). {\it Right:} \Ha, \SII\, Surface Brightness vs Diameter of M\,31 (\citealt{Lee2014a}).}
\end{figure}

\section{SNRs in the optical: Age - Evolutionary stage}

One of the main objectives in investigating SNRs is their evolution, which can be estimated in terms of their age. However, age is hard to be determined observationally; only for a handful of young or historical SNRs the actual age is known. On the other hand, their physical sizes can be measured and used as proxies of age. For that reason, many studies have adopted cumulative size distributions (hereafter CSDs) of SNRs (number of SNRs versus their diameter) in order to study the SNR evolutionary stages (e.g. \citealt{Long1990}; \citealt{Gordon1998}; \citealt{Dopita2010}; \citealt{Badenes2010}). These distributions are represented by power law forms which present different slopes ($\alpha$) at different evolutionary stages: $\alpha$=1, 2.5 and 3 for free expansion, Sedov-Taylor and radiative phases respectively. It is expected that the more evolved SNRs become, the flatter the slope gets.\\
As an example, we quote in Fig.\,4 (left plot) the SNR CSDs of the SNR populations in five well-studied galaxies: M\,31, M\,33, the Magellanic Clouds and the Milky Way (\citealt{Lee2014a, Lee2014b}). It seems that SNRs in M\,31 and M\,33 are going though the Sedov-Taylor phase, SNRs in the MCs are in the free expansion phase while SNRs in our Galaxy seem to be more evolved (radiative phase). However, these distributions should be interpreted with caution since the slopes may not only depict the evolutionary stage of the SNRs but also selection effects. For example the density distribution of the ISM may play an important role on the evolution of the SNR (e.g. \citealt{Badenes2010}) since the transition size between phases is expected to depend on the density of the surrounding ISM (along with the mass of the ejected material and the energy released by the supernova). Therefore, a dense environment would force a SNR to evolve more rapidly. Furthemore, various selection effects could influence such distributions (e.g. \citealt{Mills1984}); For example low resolution studies could give incomplete SNR samples, or different wavebands could lead to measuring different sizes for an SNR. \\
The $\Sigma$ - D (Surface Brightness vs Diameter) relation is another way of probing the evolutionary stages of SNRs. This distance-independent relation has shown that there is a slight trend of the relatively small diameter (young) SNRs to have higher surface brightnesses (e.g. Fig.\,4, right panels). What also needs to be noticed in that plots is the lack of very young objects ($<$20-30 pc), and the existence of SNRs with diameters greater than 100 pc (not typical sizes for SNRs), denoting possible misidentification of superbubbles as SNRs.

\begin{figure}
\vspace{-70pt}
\hspace{-70pt}
\begin{minipage}{.33\textwidth}
\hspace{25pt}
  \includegraphics[width=1.2\linewidth]{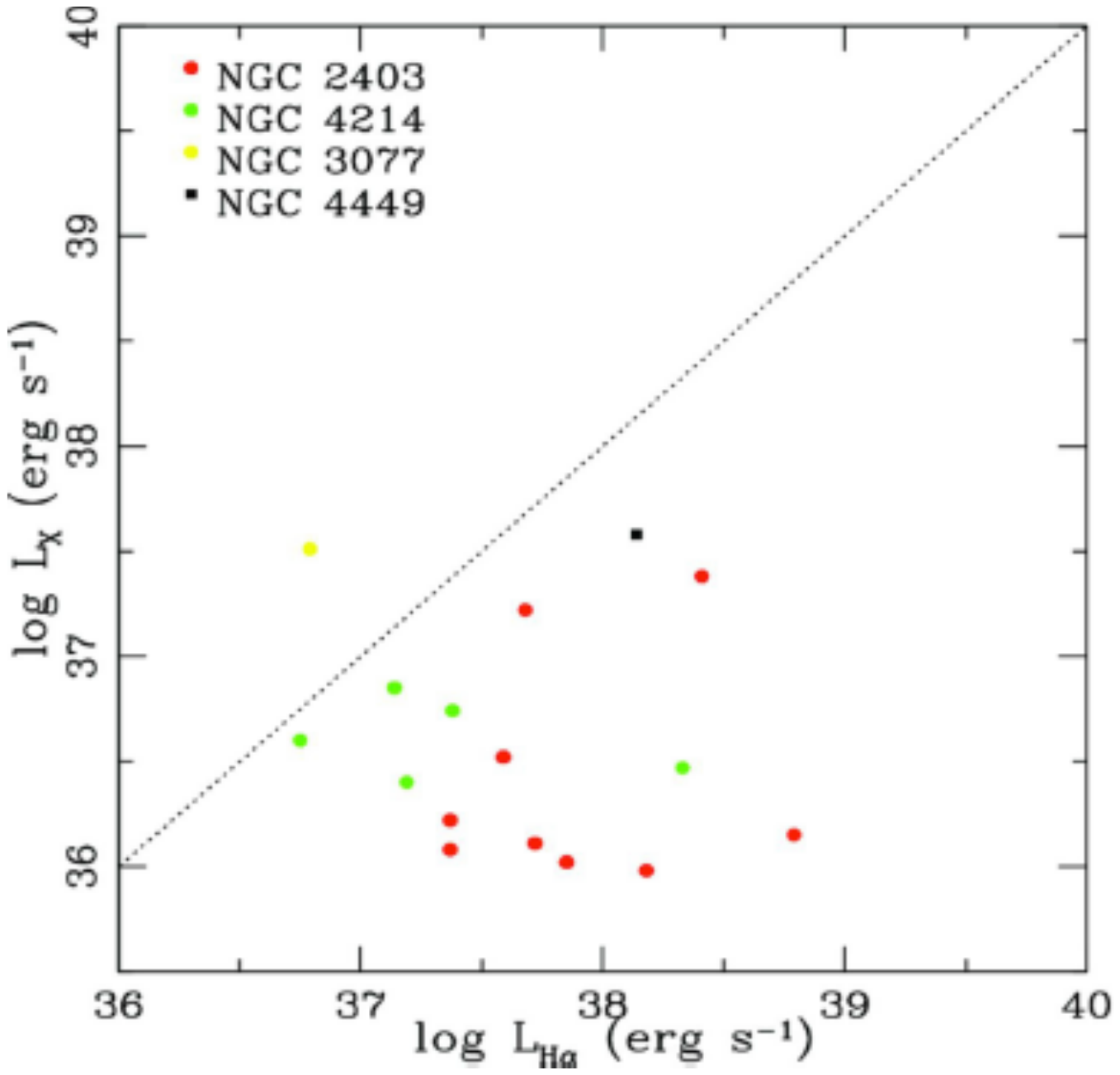}\vspace{-3pt}
\end{minipage}%
\begin{minipage}{.33\textwidth}
\hspace{30pt}
\includegraphics[width=1.03\linewidth]{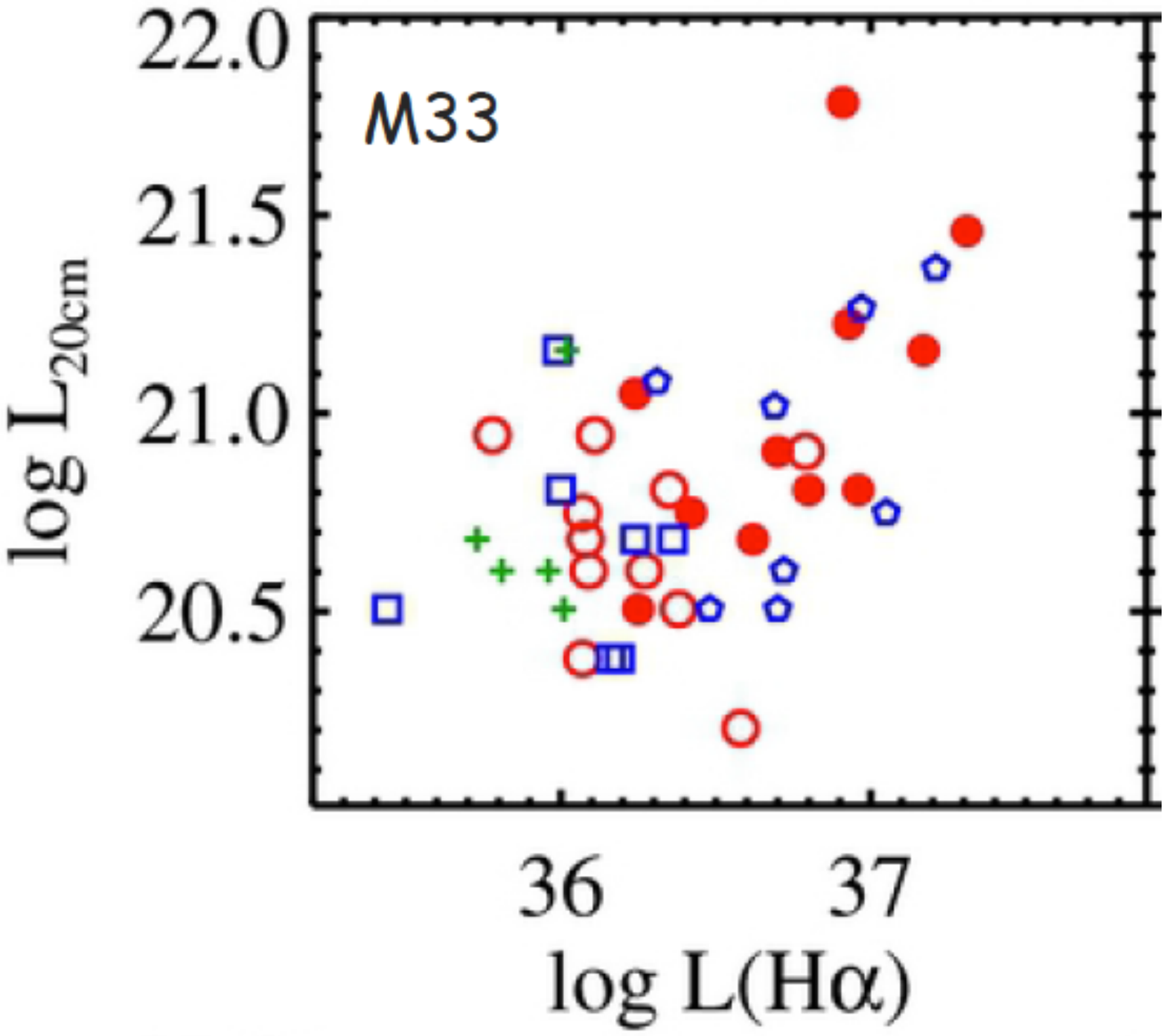}\vspace{-24pt}
\end{minipage}
\begin{minipage}{.33\textwidth}
\hspace{25pt}
  \includegraphics[width=1.0\linewidth, angle=-90]{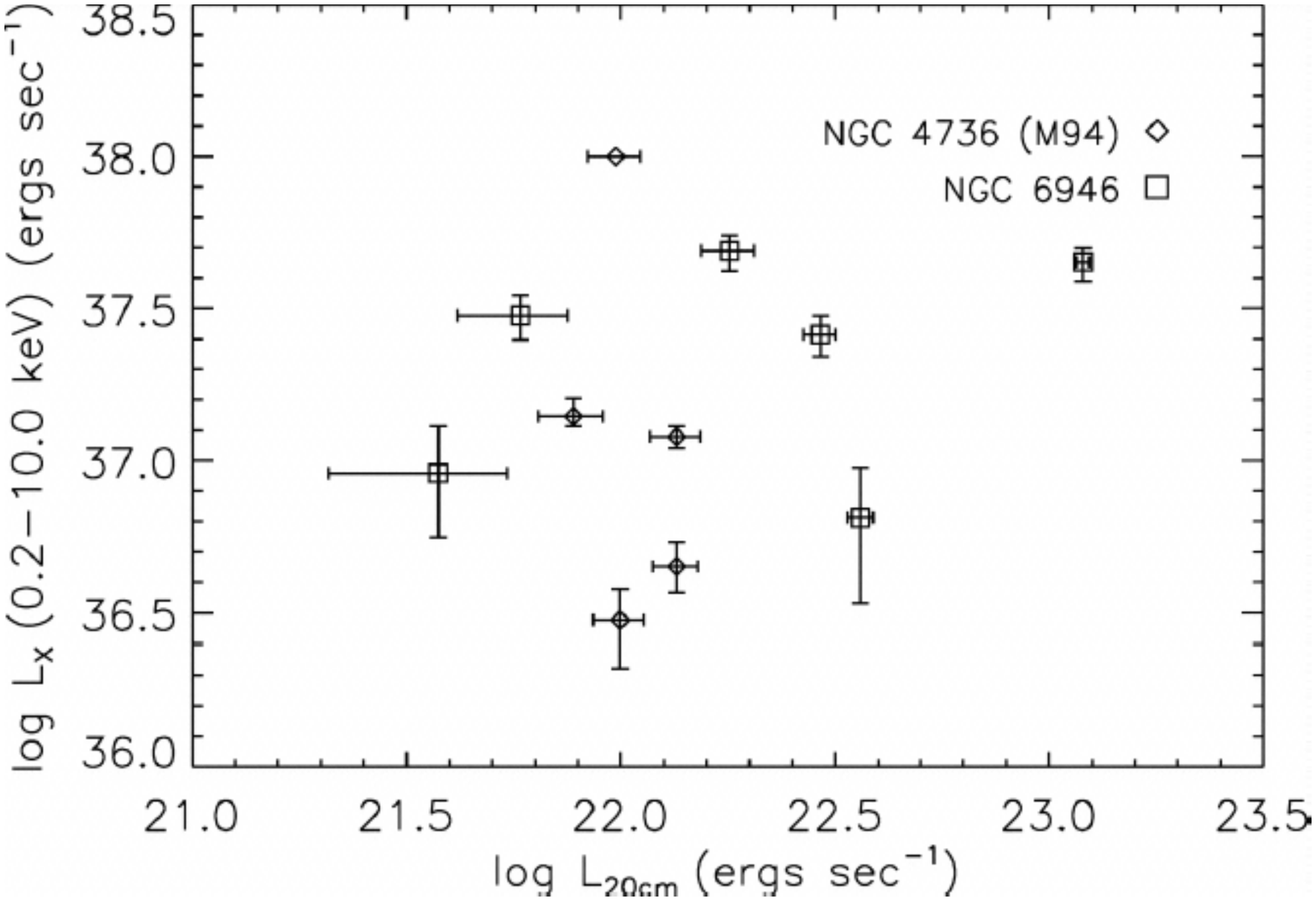}\vspace{-10pt}
\end{minipage}
\vspace{-50pt}
\caption{Multi-wavelenght luminosity relations of SNRs (see \S4.1). {\it Left: }$L_{H_a}$ - $L_{X}$ (\citealt{Leonidaki2013}). The dashed line indicates the 1:1 relation between the two luminosities. {\it Middle:} $L_{H_a}$ - $L_{radio}$. Different colours correspond to different morphological SNR  types (e.g. full/partial shells or compact; see \citealt{Lee2014b})  {\it Right: }$L_{radio}$ - $L_{X}$ (\citealt{Pannuti2007}). }
\vspace{-5pt}
\end{figure}

{\it Superbubbles}

One of the things that may impede the detection of extragalactic SNRs is their misidentification with superbubbles (cavities of hot gas in the ISM, coming either from multiple supernovae or/and blown out stellar winds of massive stars in OB associations). The shock-excited structure of these objects can grant them with moderate [S II]/H$\alpha$ values (0.45 $<$ [SII]/H$\alpha$ $<$ 0.6) (e.g. \citealt{Lasker1977}; \citealt{Chen2000}), placing them within the range of the [S II]/H$\alpha$ ratios of SNRs. The only way to discern these objects is to use their typically larger sizes ($>$100 pc), which are rare among known SNRs (e.g. \citealt{Williams1999}), and the slower \Ha\, expansion velocities ($<$100 km/s; e.g. \citealt{Franchetti2012}) than those of SNRs. On the other hand, their low-density environment is responsible for their rather faint X-ray emission (below that of SNRs: 10$^{34}$ $-$ 10$^{36}$ erg/s)  (e.g. \citealt{Chu1990}) while in the radio, they exhibit mainly thermal emission (in contrast to the synchrotron emission seen in radio SNRs).

{\it Progenitors}

The identification of the SNR progenitors (Type Ia/II) attracts the interest of several studies since the type of the progenitor can modify the SNR evolution and its interplay with the ISM. Criteria for disentangling the progenitor types are the following: a) Distinct types of objects (e.g. PWN) or objects with characteristic spectroscopic signatures (oxygen-rich, Balmer dominated) can be apparently classified as core-collapse SNRs, b) Typa Ia SNRs present statistically a more spherical mirror-symmetric morphology in X-rays than core-collapse SNRs (\citealt{Lopez2009, Lopez2011}), c) OB associations strengthen the existence of core-collapse SNRs (\citealt{Maggi2016}; \citealt{Franchetti2012}), d) Type Ia SNRs present relatively low \Ha\, fluxes compared to Type II SNRs since they tend to be located in more isolated regions (e.g. \citealt{Franchetti2012}), e) Metal abundances: Fe-rich correspond to Type Ia while O-rich correspond to Type II SNRs (\citealt{Hughes1995}; \citealt{Maggi2016}) and Fe K$\alpha$ line energy centroids:  6.4 keV for Type Ia, 6.7 keV for Type II (\citealt{Yamaguchi2014}) and f) light echoes seem to have been quite useful on classifying the progenitor type of historical SNe, based on the acquisition of their scattered-light spectrum (\citealt{Rest2005, Rest2008}).\\

\hspace{-30pt}Taking advantage the knowledge of the SNRs' ancestry, we can yield more information about their evolution and their ambient medium. For example, the left panel of Fig.\,5 shows the CSDs of all (black symbols), Type Ia (red symbols) and Type II (blue symbols) SNRs in M\,33 (\citealt{Lee2014b}). The distributions are differentiated: the number of Type Ia SNRs is smaller than Type II SNRs while the mean diameter of Type Ia remnants is larger than that of the CC remnants. This means that the majority of CC remnants may be embedded in denser ambient ISM (and hence evolve faster) than Type Ia remnants. Similarly, the surface brightness of Type Ia SNRs show a stronger linear correlation with their sizes than CC SNRs at the $\Sigma$ - D relation (e.g. Fig.\,5, right panel), indicating a less dense ISM around Type Ia SNRs.

\begin{figure}
\vspace{-40pt}
\hspace{+20pt}
\begin{minipage}{.45\textwidth}
\centering
  \includegraphics[width=0.8\linewidth]{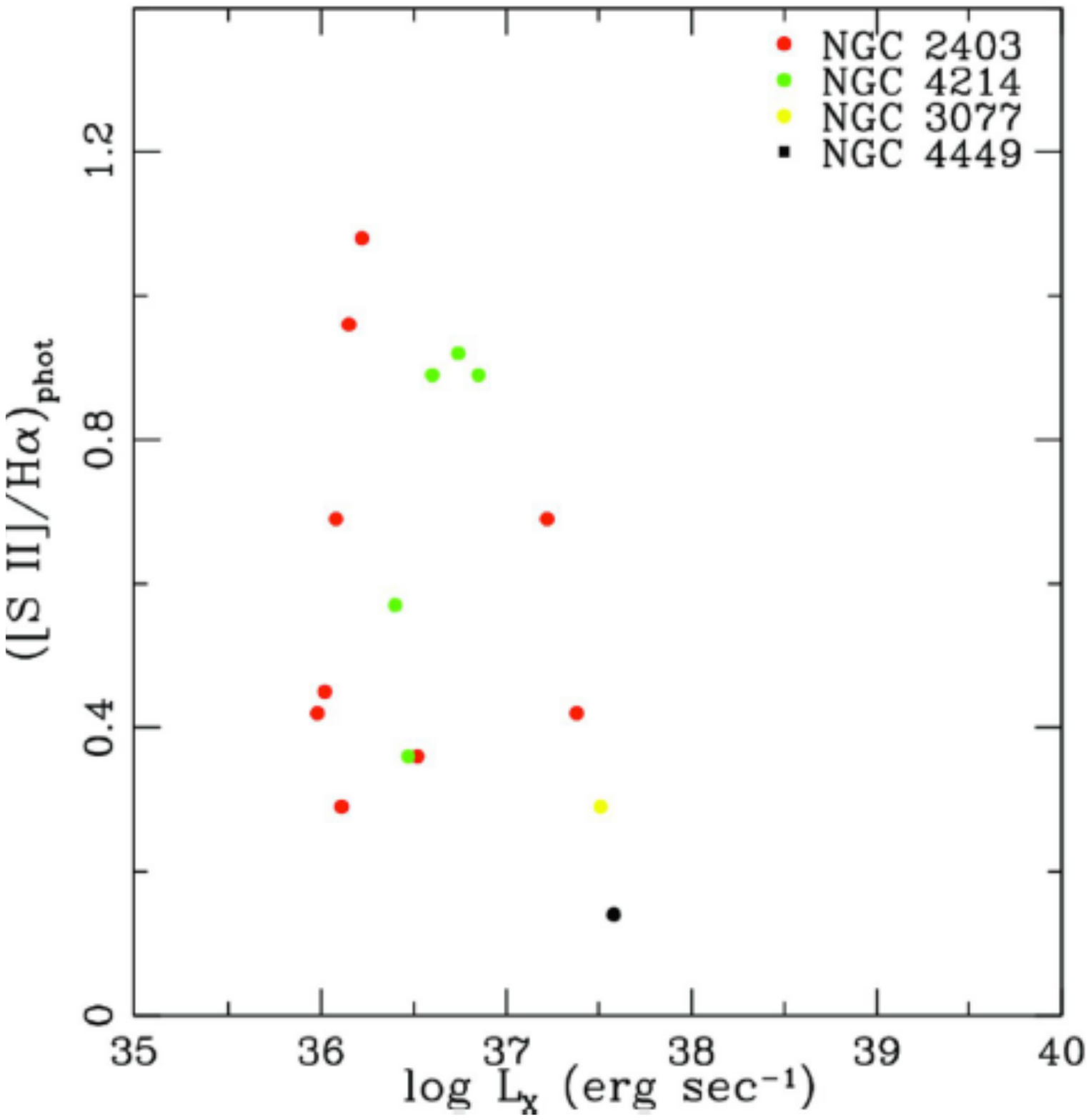}\vspace{-10pt}
  \label{fig:test1}
\end{minipage}%
\begin{minipage}{.45\textwidth}
\centering
\includegraphics[width=.75\linewidth]{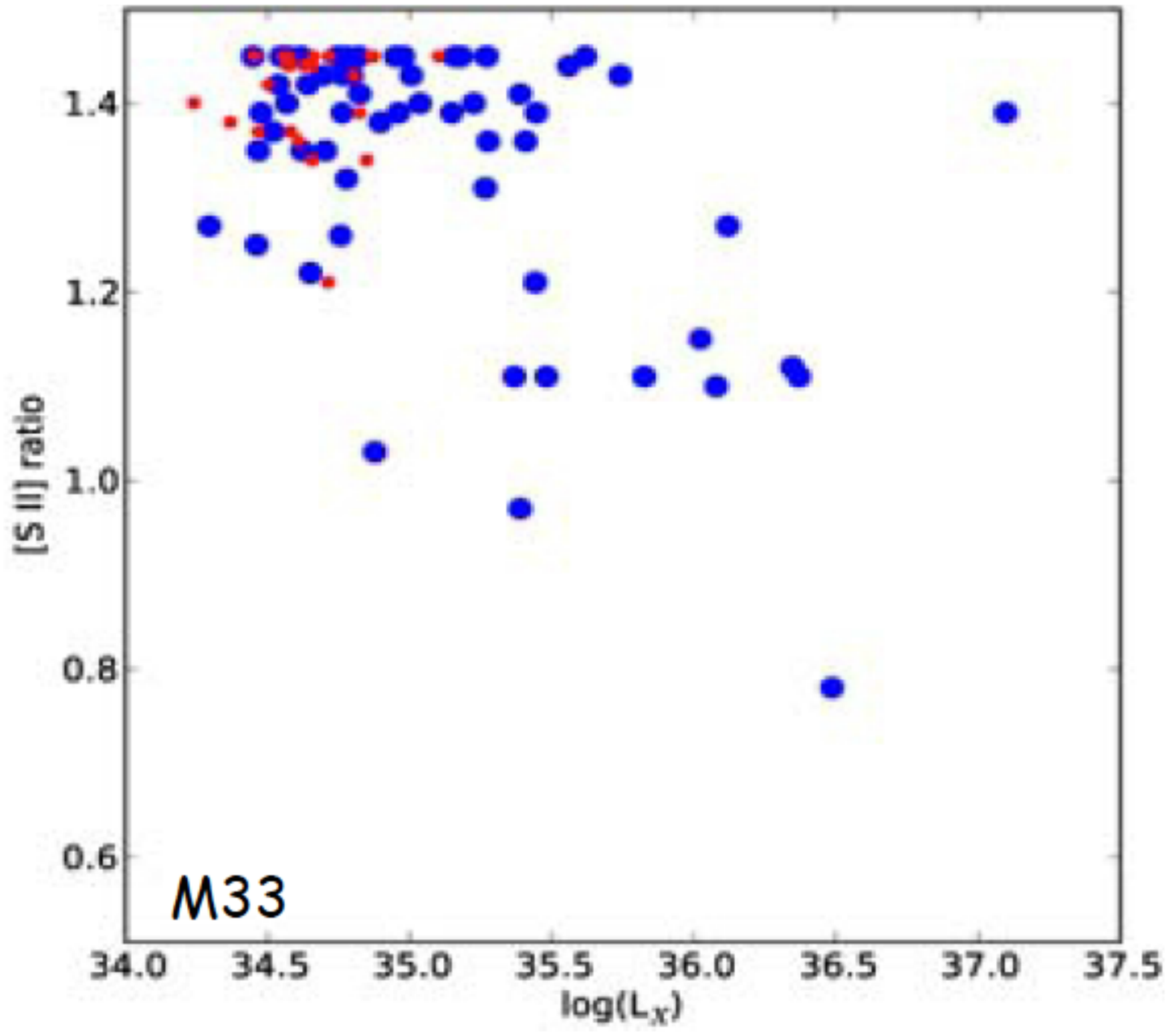}\vspace{-5pt}
\vspace{-7pt}
  \label{fig:test2}
\end{minipage}
\vspace{-35pt}
\caption{{\it Left: }$L_{X}$ - \SII/\Ha\, (\citealt{Leonidaki2013}). {\it Right: }\SII(6716\AA)/\SII(6731\AA) against $L_{X}$ (Long et al. 2010). Ratios of 1.4, 1.2, 1.0, and 0.8 correspond to densities of 50, 250, 700, and 1400 cm$^{-3}$, respectively. }
\end{figure}

\section{Multi-wavelength properties of SNRs}
\subsection{Luminosity relations}

Studies on various galaxies have shown that there is no strong correlation between the \Ha\, and X-ray luminosities of SNRs (e.g. \citealt{Pannuti2007}; \citealt{Long2010}; \citealt{Leonidaki2013}). However, most luminous X-ray SNRs tend to be the SNRs with the higher \Ha\, luminosities but with the X-ray luminosities being lower than the \Ha\, luminosities (e.g. Fig.\,6, left plot). This lack of correlation could be interpreted either as having different materials in a wide range of temperatures (\citealt{Long2010}; \citealt{Leonidaki2013}) or inhomogeneous local ISM around SNRs (e.g. \citealt{Pannuti2007}). In the same way, lack of correlation seems also to be the case between $L_{H_a}$ - $L_{radio}$ and $L_{radio}$ - $L_{X}$ of SNRs (Fig.\,6, middle and right plots).\\ 
With the same rationale we can interpret the lack of a significant correlation between the \SII/\Ha\, ratios of optically selected, X-ray emitting SNRs and their X-ray luminosities (Fig.\,7, left plot). In a sample model one would expect that stronger shocks (higher \SII/\Ha\, ratios) would correlate with higher $L_{X}$. However, because of the long-cooling time of the X-ray material, the shock velocity we are measuring does not necessarily correspond to the shock that generated the bulk of the X-ray emission material.\\
As regularly noted, density of the ambient ISM plays an important role on the morphology and the evolution of an SNR. Since higher density is a good predictor of X-ray detectability and the \SII (6717$\AA$/6731$\AA$) ratio is density sensitive, someone would expect a strong correlation between those two quanties. This can be seen in Fig.\,7 (right plot), where the \SII\, ratio against $L_{X}$ is presented for the SNRs in M\,33. In most cases the higher the density in the \SII\, zone is, the higher the X-ray luminosity is, with SNRs presenting line ratios 􏰁1.2 (density 􏰀250 cm$^{-3}$) being nearly always detected in X-rays (Long et al. 2010).

\begin{figure}
\center
\vspace{-60pt}
\includegraphics[width=10.5cm, angle=-90]{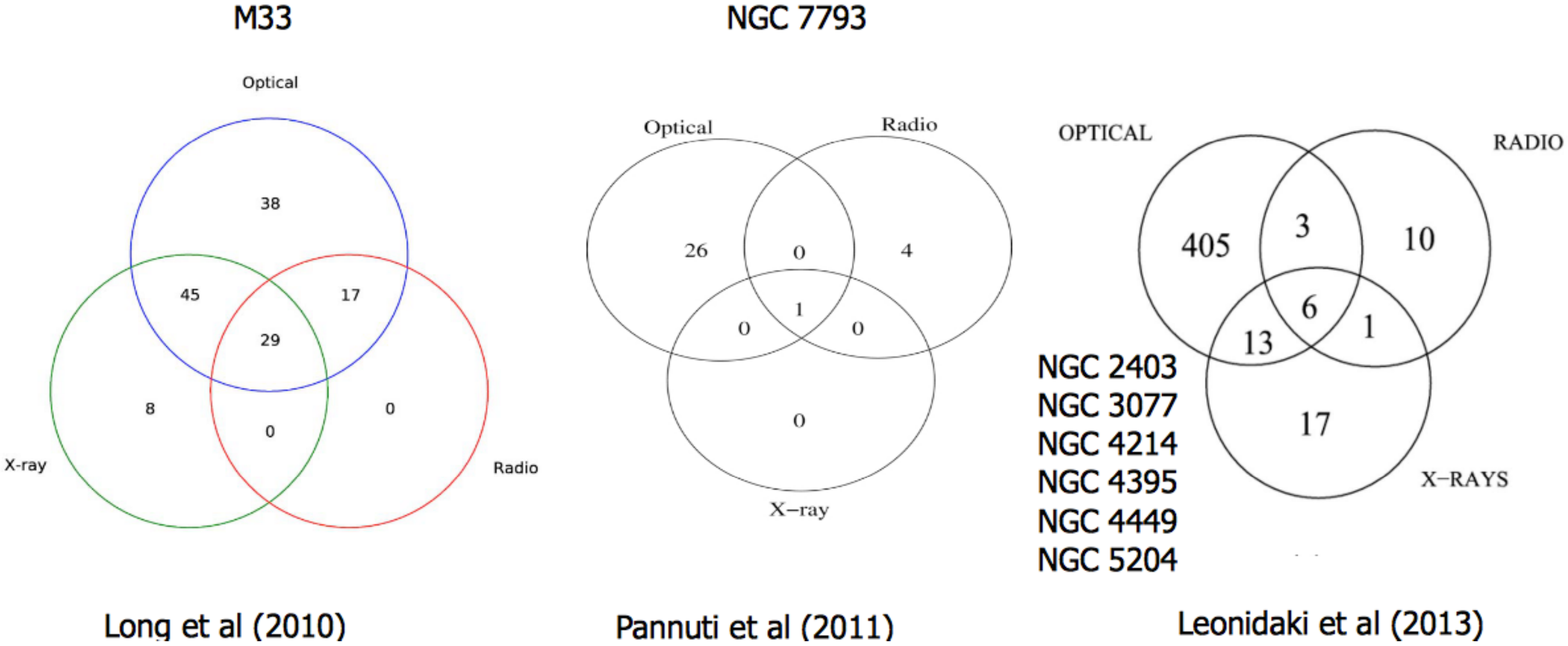} 
\vspace{-55pt}
\caption{Venn diagrams}
\end{figure}

\subsection{Venn diagrams}

Comparison of the emission of SNRs in different wavebands can provide information about the evolutionary stage of the sources and/or can illustrate selection effects. This can be shown in the form of Venn diagrams where each circle denotes a different waveband and the overlap between X-ray-, optical-, and radio-selected SNRs can be illustrated. As can be seen in Fig.\,8, there is a large gap between detection rates in different bands. In the X-rays in particular, most studies of X-ray-emitting SNRs outside the Local Group have been focused on the identification of X-ray counterparts to SNRs detected in other wavebands, rather than searching for new X-ray-emitting SNRs. For example, note the small detection rate of X-ray emitting SNRs in NGC\,7793 (Fig.\,8, middle panel) and its low overlap between other wavebands. On the other hand, when systematic studies of X-ray emitting SNR populations are performed (Fig.\,8, right panel), the detection rates increase significantly.\\
The differences in the detection rates are not only due to evolutionary effects (e.g. it's easier to detect evolved/older SNRs in the optical) but also because of the limited sensitivity of the instruments. For example, one would expect SNRs in the same type of galaxies to present similar luminosities. However there is a trend of brighter SNRs to be detected at more distant galaxies (as it is well-illustrated at Fig.\,39 from \citealt{Matonick1997}), missing the fainter SNRs. Furthermore, the detection rate of SNRs in different wavebands is strongly influenced by the properties of the surrounding medium of the source. For example, \citet{Pannuti2007} point out that optical searches are more likely to detect SNRs located in regions of low diffuse emission, while radio and X-ray searches are more likely to detect SNRs in regions of high optical confusion. We must also note that depending on the selection criteria used for detecting multi-wavelength SNRs, specific types of SNRs are excluded (e.g. Balmer-dominated/oxygen-rich in the optical, PWN in the X-rays).

\begin{figure}
\centering
\includegraphics[width=6.5cm, angle=-90]{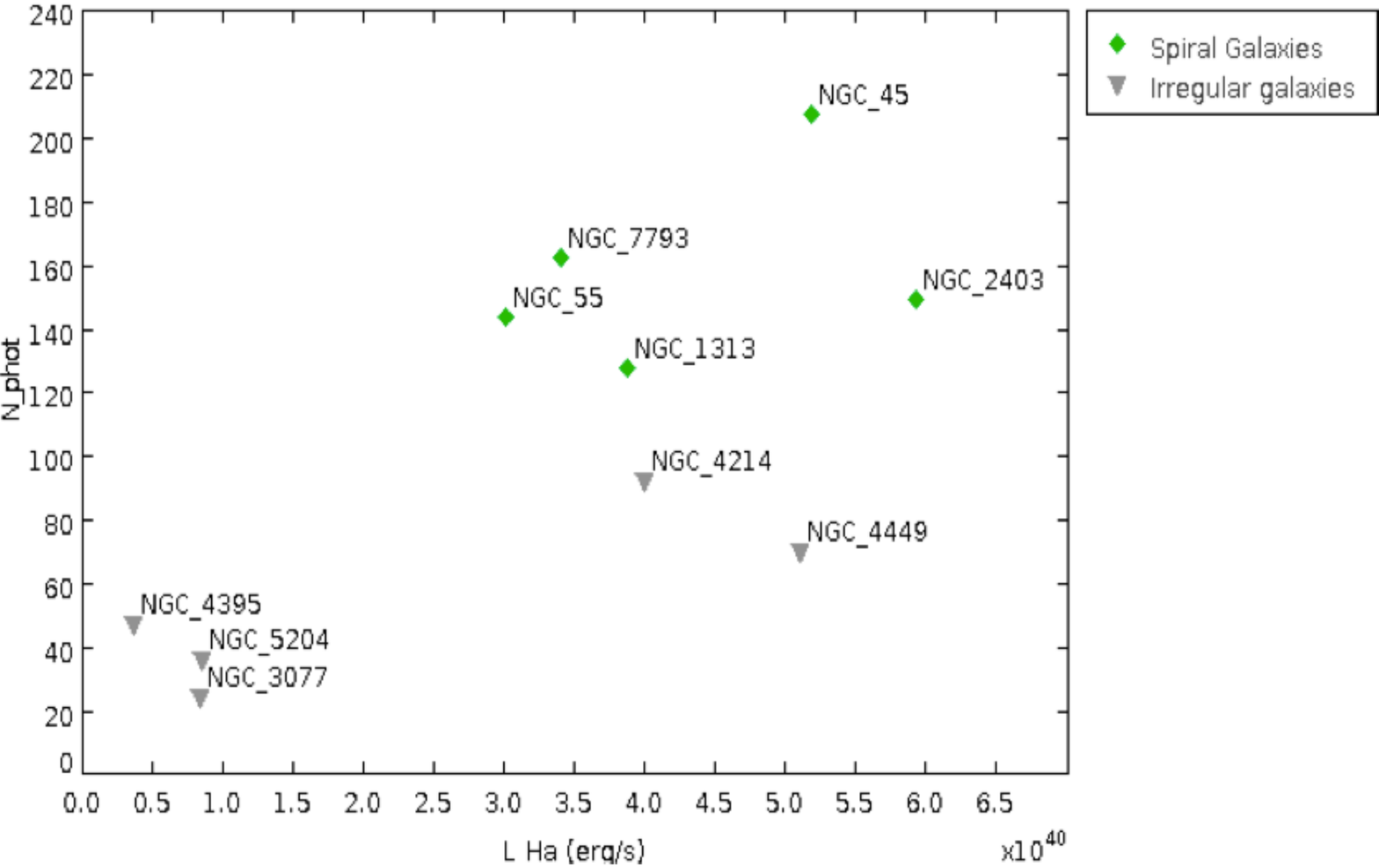} 
\caption{Number of photometric SNRs of ten nearby galaxies against the integrated \Ha\, luminosity of the host galaxy (Kopsacheili et al. in preparation).}
\end{figure}

\section{SNRs and Star Formation Rate (SFR)}

Since core-collapse SNe are the endpoints of the evolution of the most massive stars, their SNRs are good indicators of the current SFR. Therefore, a linear relation between the number of SNRs and SFRs is expected (e.g. \citealt{Leonidaki2010, Leonidaki2013}). In Fig.\,9 we show the number of photometrically observed SNRs in a sample of six nearby galaxies (NGC\,3077, NGC\,2403, NGC\,4214, NGC\,4395, NGC\,4449 and NGC\,5204; \citealt{Leonidaki2013}) against the integrated \Ha\, luminosity of each galaxy (which is used as an SFR proxy). This sample has been complemented by four more spiral galaxies (Kopsacheili et al. in preparation) in order to extend to a wider range of environments. A linear relation is apparent, with a linear correlation coefficient of 0.7 illustrating a significant correlation between the number of SNRs and SFR.

\begin{figure}
\centering
\begin{minipage}{.5\textwidth}
\centering
\includegraphics[width=.71\linewidth, angle=-90]{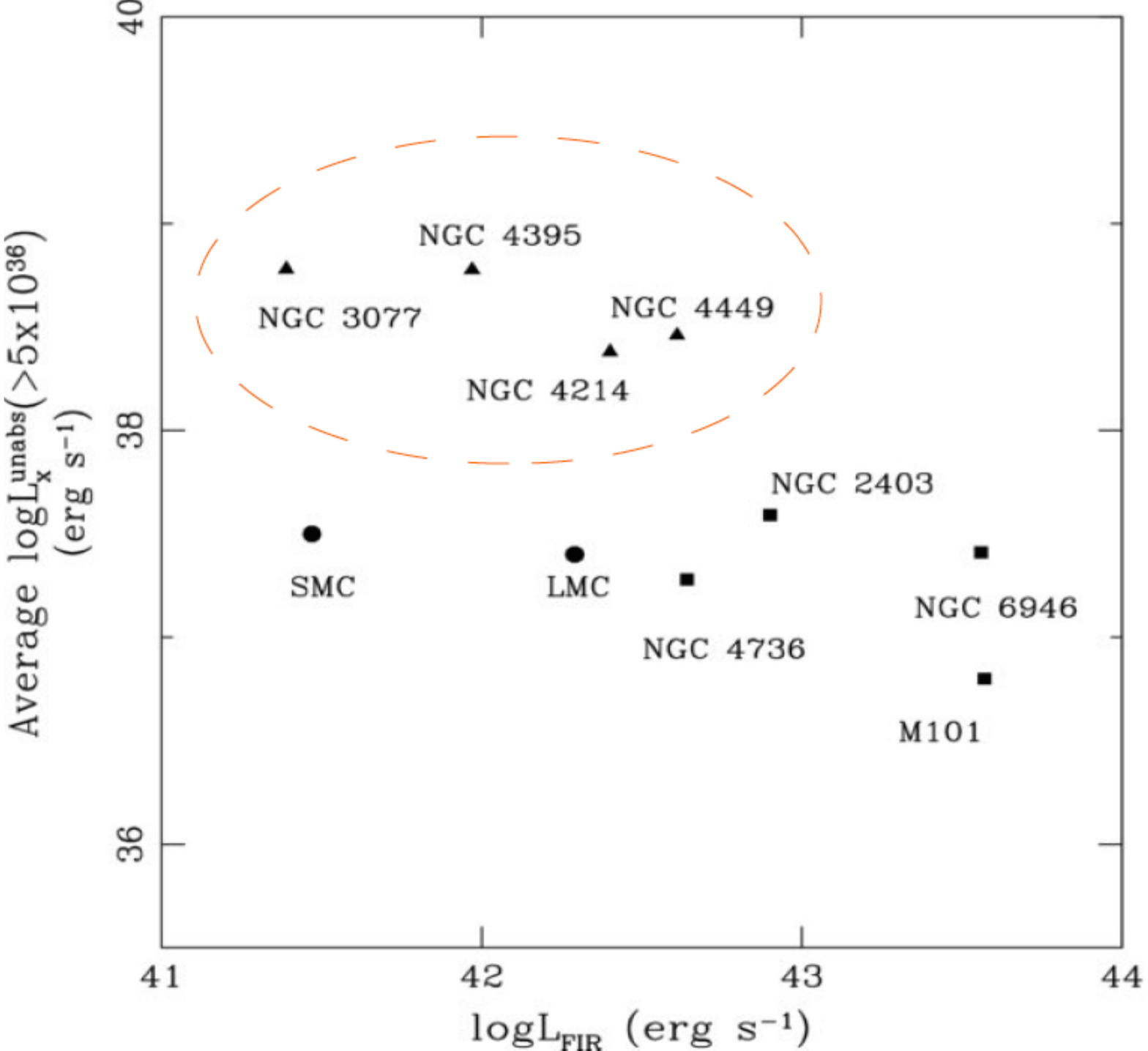}
\label{fig:test1}
\end{minipage}%
\begin{minipage}{.5\textwidth}
\centering
\vspace{+3pt}
\includegraphics[width=.72\linewidth, angle=-90]{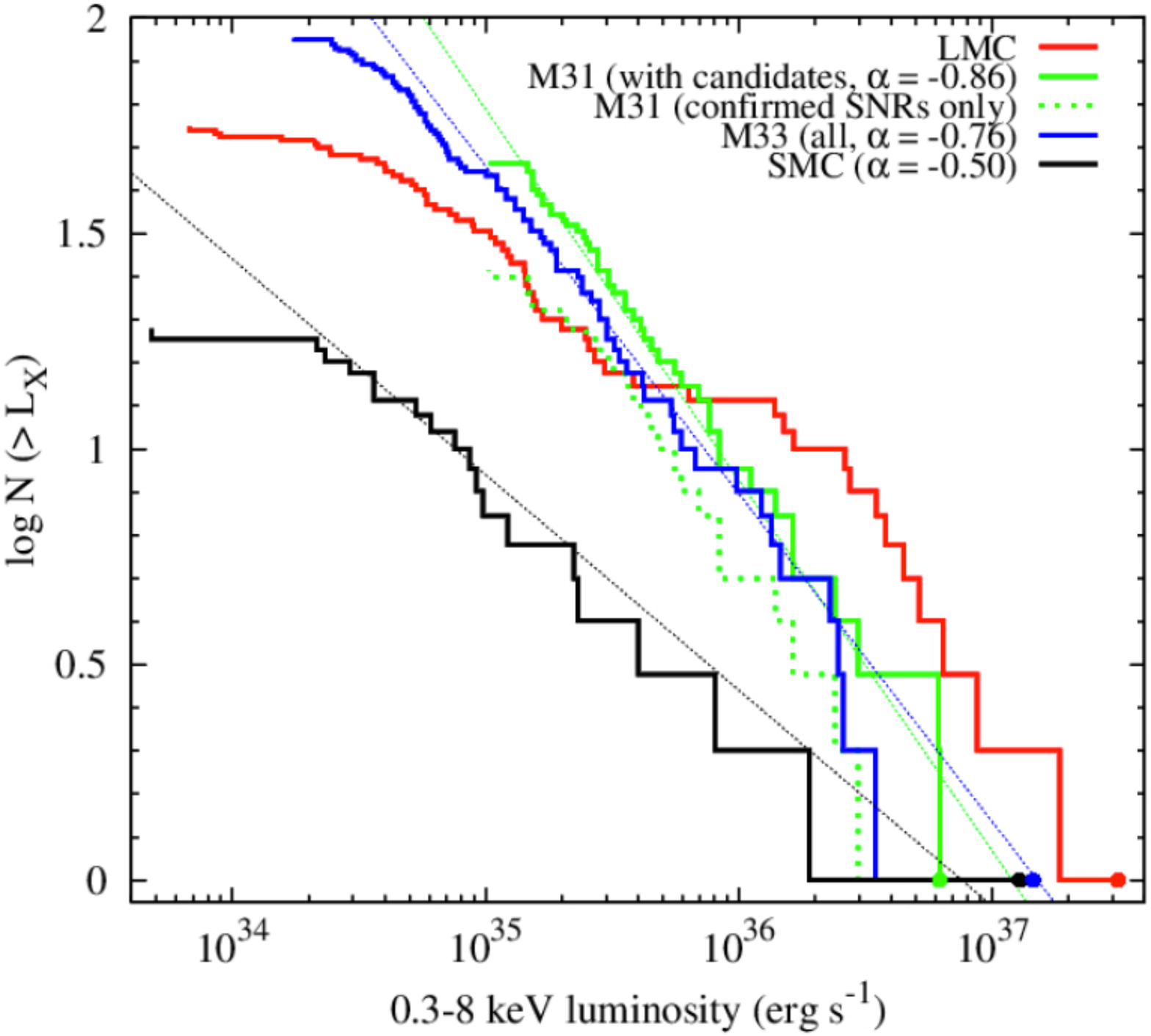}\vspace{-5pt}
\label{fig:test2}
\end{minipage}
\vspace{-5pt}
\caption{{\it Left: }Average absorption-corrected X-ray luminosity (0.3-10 keV) for SNRs in spiral (squares), irregular (triangles) galaxies and the Magellanic Clouds (circles) (\citealt{Leonidaki2010}). {\it Right: }Cumulative X-ray luminosity function of SNRs in Local Group galaxies (\citealt{Maggi2016}). See text ($\S6$) for more details.}
\end{figure}

\section{SNRs in the X-rays: Environmental effects}

Environmental effects (e.g. metallicity, ISM structure, Star Formation History) can affect the SNR populations seen in galaxies. In Fig.\,10 (left plot), the average X-ray luminosity of SNRs in various galaxies of different types is shown against the FIR luminosity of the host galaxy (\citealt{Leonidaki2010}). As expected, there is no correlation between those two quantities because the X-ray luminosity comes from individual remnants while the FIR luminosity is a property of the entire host galaxy. However, we do see a systematic trend for more luminous SNRs to be associated with irregular galaxies. This indicates a difference of the SNR population characteristics between the two samples. This could be due to the typically lower metallicity of irregular galaxies than in typical spiral galaxies (e.g. \citealt{Pagel1981}; \citealt{Garnett2002}). Low abundances result in weaker stellar winds (e.g. \citealt{Lamers1999}) which in turn produce higher mass SN progenitors. More massive progenitors are expected to produce more massive ejecta and stronger shocks which would lead to higher SNR X-ray luminosities. Another possible interpretation includes the non-uniform ISM which is often the case in irregular galaxies, where local enhancements (especially at the star-forming regions) could result to more luminous SNRs. \\
Furthermore, in the same work (\citealt{Leonidaki2010}) there was another strong indication of different SNR populations in different types of galaxies, by comparing the luminosity distributions of X-ray emitting SNRs. They found that the numbers of SNRs in irregular galaxies are more consistent with an Maggelanic Cloud-like SNR X-ray luminosity function (XLF), while those of spiral galaxies are more consistent with the SNR-XLF of the spiral M33. However, due to the small number of statistics, this result could not be quantified. However, \citet{Maggi2016} pinned the environmental impact on SNR populations down by comparing the X-ray luminosity functions of SNRs in Local Group galaxies (LMC, SMC, M31, and M33; Fig.\,10 right panel). They revealed different slopes of the XLFs which illustrate diferences between the SNR populations in these galaxies.

\section{Concluding...}


There has been a revolution on the investigation of extragalactic SNRs the last decades, enabling the study of their physical properties in different environments. However, this is just the beginning; there is an imperative need in observing more galaxies at larger depths and in a multi-wavelength context in order to alleviate the selection effects that hamper the current studies of SNRs and obtain a more complete picture of the SNR populations. \\

\small  
%

%


\end{document}